\documentclass[aps,prd,onecolumn,11pt,groupedaddress,nofootinbib,showpacs]{revtex4}
\usepackage{amsmath,amssymb}
\begin{document}

\title{The perturbative structure of spin glass field theory
}

\author{T.~Temesv\'ari}
\email{temtam@helios.elte.hu}
\affiliation{
MTA-ELTE Theoretical Physics Research Group,
E\"otv\"os University, P\'azm\'any P\'eter s\'et\'any 1/A,
H-1117 Budapest, Hungary
}

\date{\today}

\begin{abstract}
Cubic replicated field theory is used to study the glassy phase of the short-range
Ising spin glass just below the transition temperature, and for systems above, at,
and slightly below the upper critical dimension six. The order parameter function
is computed up to two-loop order. There are two, well-separated bands in the mass
spectrum, just as in mean field theory. The small mass band acts as an infrared
cutoff, whereas contributions from the large mass region can be computed
perturbatively ($d>6$), or interpreted by the $\epsilon$-expansion around the
critical fixed point ($d=6-\epsilon$). The one-loop calculation of the
(momentum-dependent) longitudinal mass, and the whole replicon sector is also
presented. The innocuous behavior of the replicon masses while crossing the
upper critical dimension shows that the ultrametric replica symmetry broken
phase remains stable below six dimensions.
\end{abstract}
\maketitle

\section{Introduction}

A spin glass is a prototype of complex systems, with its slow dynamics on
macroscopic time scales, unusual equilibrium properties, and complicated
phase space structure which breaks ergodicity. The interest in the understanding
of the spin glass problem started in the seventies of the last century,
and lots of results have accumulated since then, nevertheless many basic
questions have remained open. (For an overview of the history of spin glass research,
see review papers from different periods: \cite{BinderYoung,MePaVi,FischerHertz,%
SG98}.

Numerical simulations are important tools for getting information about spin
glass properties. Without trying to overview this huge field, we only mention here
the Janus Collaboration using the special purpose Janus computer, providing
results about the spin glass phase in the physical three-dimensional
Edwards-Anderson model which are compatible with an ultrametrically organized
replica symmetry broken (RSB) phase (see \cite{Maiorano_et_al} for a recent
list of references related to the Janus Collaboration). This ultrametric glassy
phase emerged for the first time in the solution of the mean field theory of
the Ising spin glass by Parisi, see \cite{MePaVi} and references therein, and
--- according to our present knowledge --- it seems to persist below
the upper critical dimension six \cite{PT};
possibly (as the aforementioned
numerical simulations suggest) down to three dimensions. Nevertheless the
details of the RSB phase of the short range finite-dimensional model differ in
many ways from its mean field counterparts. Important examples for such
discrepancies are the leading behavior of the order parameter
function and momentum-dependent masses (or, equivalently, correlation functions)
close to criticality. Moreover, these details depend on the space dimension $d$
which can be well illustrated by the breakpoint $x_1$ of the order parameter
function $q(x)$, see Refs.\ \cite{old_1991,beyond,scaling_and_infrared,PT}:
$x_1$ is proportional to $\tau\sim (T_c-T)/T_c$ in mean field
theory (which is equivalent to the infinite-dimensional model), and this behavior
persists down to $d=8$, with possibly a logarithm of $\tau$ at exactly eight
dimensions. For $6<d<8$, $x_1\sim \tau^{\frac{d}{2}-3}$, whereas at exactly six
dimensions: $x_1\sim |\ln \tau |^{-1}$. Below six dimensions $x_1$ becomes finite
at $T_c$, and renormalization group arguments show \cite{PT} that its critical
value is universal. It was computed in first order in $\epsilon=6-d$ in
Ref.\ \cite{PT}, the present paper extends this calculation to second order,
see Eq.\ (\ref{x_1_below_2}). There is a trend of increasing $x_1$ with
decreasing $d$, a clear sign that RSB becomes more dominant. This contradicts
expectations that a replica symmetric (RS) glassy phase, which is characterized
by the so-called ``droplet'' picture \cite{FiHu86,BrMo86,FiHu88}, enters in
some low dimension, for which a possible scenario would be if $x_1$ decreased to zero.
Well below $d=6$ one expects $x_1$ to be of order unity,
meaning that intervalley overlaps become as important as self-overlap
\cite{MePaVi}.

As an alternative to numerical simulations, replicated field theory provides
analytic results, and it has the advantage that space dimension can be
chosen at will by defining the model on a $d$-dimensional hypercubic
lattice. In the present paper, spin glass field theory is studied below eight
dimensions, also passing through the upper critical dimension six. (In this
domain of dimensions, a simple cubic model with the coupling constant $w$ is
sufficient for obtaining critically relevant properties, the quartic coupling
of the truncated model, for instance, which is dangerously irrelevant in higher
dimensions,
can now be neglected.) Our main purpose is to understand how the perturbative method
works in this system whose mass spectrum consists of two separated bands:
a large one dominating the behavior in the ``near infrared''
momentum range, and a small one
--- extending to zero --- related to the ``far infrared'' sector. 
We extend former calculations of the order parameter function $q(x)$ to two-loop
order. This calculation needs computing some of the one-loop self-energy
insertions; the results can be used to get the momentum-dependent longitudinal
mass and the replicon band in one-loop order. The findings for the replicon band
support the idea that stability of the RSB phase persists below six dimensions.

The paper is organized as follows: The model is defined and the equation of state
for the order parameter function presented in Sec.\ \ref{model}, while the properties
of the free propagators are studied in Sec.\ \ref{free_propagator}.
The different terms contributing to the equation of state up to two-loop order
are worked out in Sec.\ \ref{contributions}.
Section \ref{results} contains the main results for the order parameter function
in the different dimensional regimes, namely $6<d<8$, $d=6$, and $d\lesssim 6$.
This section is divided into three subsections: \ref{result_x1} is for the breakpoint
$x_1$, \ref{result_q1} for the Edwards-Anderson order parameter,
while subsection \ref{result_delta_q} is devoted to $q(x)$ with $x<x_1$.
The momentum-dependent mass (the inverse propagator) is studied in Sec.\ \ref{mass}.
In subsection \ref{exp2} the longitudinal mass is computed, and its scaling
behavior below six dimensions displayed and proved. The replicon sector is left
to \ref{replicon}. The discussion of the results is included in the last section,
i.e.\ in Sec.\ \ref{conclusion}.
Several results are summarized in listed and tabulated forms
in the three appendices.

\section{The replicated cubic field theory and mass renormalization}
\label{model}

The replicated field theory (representing the Ising spin glass on a $d$-dimensional
hypercubic lattice in zero external magnetic field) has its dynamical variables
$\phi^{\alpha\beta}=\phi^{\beta\alpha}$ --- with $\phi^{\alpha\alpha}=0$
and replica indices $\alpha,\beta,\dots$ taking values from $1$ to $n$ where $n$ is
the replica number --- and Lagrangian $\mathcal L(\phi^{\alpha\beta})$ which is
invariant under any permutations of the replicas (a trivial outcome of the replica
trick) {\em and} under the transformation $\phi^{'\alpha\beta}=
(-1)^{\alpha+\beta}\,\phi^{\alpha\beta}$ (expressing the extra symmetry coming with
the vanishing external field \cite{droplet}).
Concentrating on the glassy phase just below the critical temperature, invariants
of the above symmetry which are higher than cubic can be neglected
when $d<8$, and
the Lagrangian takes the relatively simple form
\begin{equation}\label{simple_L}
\mathcal{L}=
\frac{1}{2}\sum_{\mathbf p}
 \bigg(\frac{1}{2} p^2+ m_c-\tau\bigg)\sum_{\alpha\beta}
\phi^{\alpha\beta}_{\mathbf p}\phi^{\alpha\beta}_{-\mathbf p}
-\frac{1}{6N^{1/2}}\,\,w\,\sideset{}{'}\sum_{\mathbf {p_1p_2p_3}}
\sum_{\alpha\beta\gamma}\phi^{\alpha\beta}_{\mathbf p_1}
\phi^{\beta\gamma}_{\mathbf p_2}\phi^{\gamma\alpha}_{\mathbf p_3}\quad.
\end{equation}
Momentum conservation is indicated by the primed
summation.
The number $N$ of the Ising spins becomes infinite in the thermodynamic
limit, rendering summations to integrals over the continuum of momenta
in the diagrams of the perturbative expansion. A momentum cutoff $\Lambda$
is always understood to block ultraviolet divergences.
The two important parameters of the model are the reduced temperature $\tau$
and the coupling constant $w$. As we are working in the immediate vicinity
of the critical point, $\tau$ is assumed to be much smaller than $\Lambda^{2}$.
The critical bare mass can be computed relatively easily
yielding (up to second order and for $n=0$):
\begin{multline}\label{m_c}
m_c=m_c^{(1)}+m_c^{(2)}=\\[3pt]
=-w^2\frac{1}{N}\sum_{\mathbf p}\frac{1}{p^4}
+w^4\frac{1}{N^2}\sum_{\mathbf p,\mathbf q} \left\{ \frac{4}{p^6q^2}\left[
\frac{1}{(\mathbf p-\mathbf q)^2}-\frac{1}{q^2}\right]
+\frac{1}{p^4q^2(\mathbf p-\mathbf q)^2}\left[\frac{1}{q^2}+
\frac{1}{(\mathbf p-\mathbf q)^2}\right]\right\} \quad.
\end{multline}
For studying the glassy phase, the replica symmetric Lagrangian above is
converted to the replica symmetry broken one by the transformation
$\phi^{\alpha\beta}_{\mathbf p} \longrightarrow \phi^{\alpha\beta}_{\mathbf p}
-\sqrt N\,q_{\alpha\beta}\,\delta^{\text{Kr}}_{\mathbf p=\mathbf 0}$ where 
$q_{\alpha\beta}\equiv \langle \phi^{\alpha\beta}_{i}\rangle$
is the {\em exact} homogeneous order parameter matrix.
The Lagrangian in (\ref{simple_L}) gets then the additional mass term
$-\frac{1}{2}w\sum_{\mathbf p}\sum_{\alpha\beta\gamma}q_{\alpha\beta}\,
\phi^{\beta\gamma}_{\mathbf p}\phi^{\gamma\alpha}_{-\mathbf p}$.
The new fields have, by definition, zero mean now, and this condition yields the
equation of state for $q_{\alpha\beta}$:
\begin{equation}\label{eq_of_state}
2\tau\, q_{\alpha\beta}+w\,(q^2)_{\alpha\beta}+w\frac{1}{N}\sum_{\mathbf p}
\sum_{\gamma\not=\alpha,\beta}G^{\text{exact}}_{\alpha\gamma,\beta\gamma}
-2m_c\,q_{\alpha\beta}=0\quad,
\end{equation}
where the exact propagator satisfies Dyson's equation:
\begin{equation}\label{Gamma}
(G^{\text{exact}})^{-1}=p^2+M-\Sigma \quad \text{with} \quad
\left\{
\begin{aligned}
M_{\alpha\beta,\alpha\beta} &=-2\tau+2m_c\\
M_{\alpha\gamma,\beta\gamma}&=-wq_{\alpha\beta}\\
M_{\alpha\beta,\gamma\delta}&=0
\end{aligned}
\right.
 \qquad \text{and} \quad \Sigma \quad \text{the self-energy}.
\end{equation}
From now on, $n$ is set to zero (the spin glass limit), 
and an infinite-step ultrametric
structure \cite{MePaVi} is assumed
for $q_{\alpha\beta}=q(x)$, with the overlap $x=\alpha\cap\beta$.
Construction of the free propagator $G$ in spin glass field theory is not a trivial
task. One can be guided by lessons learnt from
Refs.\ \cite{old_1991,beyond,scaling_and_infrared}: $\tau=wq_1+\dots$ and $q(x)\sim
x$ in leading order. To generate a perturbation theory where the small parameter
is the breakpoint $x_1$ of the order parameter function, one can divide the mass
as $M=M_0+M_1$, and $q$ as 
$q(x)=q_1\bar q(r)=q_1[r+\Delta \bar q(r)]$ with $r\equiv x/x_1$.
By definition, $\Delta \bar q(1)=0$ --- since $q_1$ and $x_1$ are exact quantities
---, and for the free propagator we have $G^{-1}\equiv p^2+M_0$ with
\begin{equation}\label{M_0}
\begin{aligned}
(M_0)_{\alpha\beta,\alpha\beta} &=-2wq_1\,(1-x_1/2)\\
(M_0)_{\alpha\gamma,\beta\gamma}&=-wq_1\, r,\quad\text{with}\quad r=x/x_1
\quad\text{and}\quad x=\alpha\cap\beta\\
(M_0)_{\alpha\beta,\gamma\delta}&=0\quad.
\end{aligned}
\end{equation}
$M_1$, on the other hand, plays the role of a quadratic counter term, and the exact
propagator has the following expansion:
\begin{equation}\label{G_exact}
G^{\text{exact}}=G+G\,(\Sigma -M_1)\,G+\dots
\end{equation}
with
\begin{equation}\label{M_1}
(M_1)_{\alpha\beta,\alpha\beta}=-2\tau+2wq_1+2m_c-x_1wq_1\equiv \delta M\qquad
\text{and}\qquad (M_1)_{\alpha\gamma,\beta\gamma}=-wq_1\,\Delta \bar q(r)\quad.
\end{equation}

\section{The free propagator}\label{free_propagator}

One can follow Ref.\cite{beyond} step by step to construct the free propagator from
$M_0$ in Eq.(\ref{M_0}). The replicon masses are easily found and can be displayed
in the usual parametrization as
\begin{equation}\label{free_replicon}
\lambda_0(x;u,v)=\frac{1}{2}x_1wq_1\,
\Big[\Big(\frac{u}{x_1}\Big)^2+
\Big(\frac{v}{x_1}\Big)^2\Big],\qquad 0\le x \le u,v \le x_1\quad.
\end{equation}
Due to the second term of the diagonal element in Eq.\ (\ref{M_0}), this replicon
band extends from zero to $x_1wq_1$, and 
it is necessary for having a positive mass spectrum.
While the replicon eigenvalues are exact and of order $\sim x_1wq_1$,
the band of large masses is centered around $2wq_1$ and has the expansion:
\[
\lambda_0(k)=2wq_1\Big\{1+\frac{1}{6}\Big[2\Big(\frac{k}{x_1}\Big)^3-1\Big]\,x_1
+O(x_1^2)\Big\}
,\qquad 0\le k\le x_1\quad.
\]
One can see that the small parameter $x_1$ has a double role in the mass 
spectrum: Firstly, it is the ratio of the small to large masses which are then
clearly separated and, secondly, the widths of both bands are
proportional to $x_1$.

Inversion of the mass operator is complicated, but 
feasible by the techniques of Ref.\ \cite{beyond}. As an illustration of the
behavior of the free propagator components as a function of the momentum,
let us consider the combination entering the equation of state
(\ref{eq_of_state})%
\footnote{For the parametrization of the components of an ultrametric matrix,
see Ref.\ \cite{beyond}.\label{parametrization}}:
\begin{equation}\label{Y}
Y_{\alpha\beta}\equiv Y(x)\equiv 
\sum_{\gamma\not=\alpha,\beta}G_{\alpha\gamma,\beta\gamma}=
-\int_0^x dy\, G^{yy}_{1x} -x\, G^{xx}_{\,\,1}-2\int_x^1dy G^{xy}_{\,\,1}
,\qquad x=\alpha\cap\beta\quad.
\end{equation}
$Y$ has the following expansion in the large (l) and small (s) mass regimes:
\begin{itemize}\item
Large mass regime, i.e.\ $p^2\sim 2wq_1$:
\[
Y=\frac{1}{p^2}\,\Big[G^{(l)}_0(u;r)+x_1\,G^{(l)}_1(u;r)+
x_1^2\,G^{(l)}_2(u;r)\dots\Big]\quad,
\quad u=\frac{p^2}{2wq_1}\quad\text{and}\quad r=\frac{x}{x_1}\quad.
\]
When $p^2<2wq_1$, we have the following expansion of the $G^{(l)}$
functions (although not indicated, the $g^{(l)}$ coefficients are still
functions of $r$):
\begin{align*}
G^{(l)}_0(u)&=u^{-1}\,(g^{(l)}_{00}+g^{(l)}_{01}u+g^{(l)}_{02}u^2+\dots),\\
G^{(l)}_1(u)&=u^{-2}\,(g^{(l)}_{10}+g^{(l)}_{11}u+\dots),\\
G^{(l)}_2(u)&=u^{-3}\,(g^{(l)}_{20}+\dots),\quad\dots\quad.
\end{align*}
\item
Small mass regime, i.e.\ $p^2\sim x_1\times 2wq_1$:
\[
Y=\frac{1}{x_1p^2}\,\Big[G^{(s)}_0(u;r)+x_1\,G^{(s)}_1(u;r)+
x_1^2\,G^{(s)}_2(u;r)\dots\Big],
\quad u=\frac{p^2}{x_1\,2wq_1}\quad\text{and}\quad r=\frac{x}{x_1}\quad.
\]
The expansion of the $G^{(s)}$ functions for $p^2>x_1\,2wq_1$
(the $r$-dependence of the coefficients not indicated):
\begin{align*}
G^{(s)}_0(u)&=u^{-1}\,(g^{(s)}_{00}+g^{(s)}_{01}u^{-1}+g^{(s)}_{02}u^{-2}
+\dots),\\
G^{(s)}_1(u)&=(g^{(s)}_{10}+g^{(s)}_{11}u^{-1}+\dots),\\
G^{(s)}_2(u)&=u\,(g^{(s)}_{20}+\dots),\quad\dots\quad.
\end{align*}
\end{itemize}
It is $G^{(l)}_0$ which can be computed by the least effort using the
``smallest block approximation" (SBA), meaning that the $\gamma$ summation
in (\ref{Y}) is restricted to the smallest ultrametric blocks of size $x_1$
around $\alpha$ and $\beta$:
\begin{equation}\label{G_0}
Y\approx 2(x_1-1)G^{xx_1}_{\,\,\,1}\approx -2\,G^{xx_1}_{\,\,\,1}
\Longrightarrow
G^{(l)}_0(u;r)=\frac{-(u^2+3u+1)r+r^3}{u(u+1)^2}
\end{equation}
where the near infrared form of $G^{xx_1}_{\,\,\,1}$
from Sec.\ 6 of Ref.\ \cite{beyond}
has been used. The calculation of $G^{(l)}_1$ --- which is necessary for 
obtaining the
corrections to the order parameter function --- is somewhat complicated, and the result
is not displayed here 
[but Eq.\ (\ref{F_1}) is based on that
calculation].

There is complete matching in the momentum range $x_1\,2wq_1<p^2<2wq_1$ which is
satisfied by the following matching-conditions between the two types of coefficients:
\[
g^{(l)}_{00}=g^{(s)}_{00};\quad g^{(l)}_{10}=g^{(s)}_{01},\quad
g^{(l)}_{01}=g^{(s)}_{10};\quad g^{(l)}_{20}=g^{(s)}_{02},\quad
g^{(l)}_{11}=g^{(s)}_{11},\quad g^{(l)}_{02}=g^{(s)}_{20};\quad \dots\quad.
\]

\section{Contributions to the equation of state up to two-loop order}
\label{contributions}

In this section, a summary of the results for the different terms appearing
in Eq.\ (\ref{eq_of_state}) are presented.
\subsection{The zero-loop term}

Assuming $q(x)=q_1[r+\Delta \bar q(r)]$ and $r= x/x_1$,
the first two terms in (\ref{eq_of_state})
can be written (for $\alpha\cap\beta=x$) as
\begin{multline*}
2\tau\, q(x)+w\,q^2(x)=
2\tau\, q_1[r+\Delta \bar q(r)]\\+
wq_1^2\Big[-2r+x_1(r-\frac{1}{3}r^3)-2\Delta \bar q(r)+2x_1\{r;\Delta \bar q(r)\}
+x_1\{\Delta \bar q(r);\Delta \bar q(r)\}\Big]
\end{multline*}
with the bilinear form defined by
\begin{equation}\label{bilinear}
\{f(r);g(r)\}\equiv f(r)g(1)+f(1)g(r)-f(r)\int_r^1du g(u) -g(r)\int_r^1du f(u)-rf(r)g(r)
-\int_0^rdu f(u)g(u).
\end{equation}
The last term which is quadratic in $\Delta \bar q$ will be neglected in the present
stage of approximation. We will need the following derivatives:
\begin{subequations}\begin{gather}\label{first}
\frac{d}{dr}\,\big[2\tau\, q(x)+w\,q^2(x)\big]
\Bigg |_{r=1}=
q_1(2\tau-2wq_1)\big[1+{\Delta \bar q\,}'(r=1)\big],\\ \label{second}
\frac{d^2}{dr^2}\,\big[2\tau\, q(x)+w\,q^2(x)\big]=
-2x_1wq_1^2\,r-4x_1wq_1^2\,r\,{\Delta \bar q\,}'+
\big[q_1(2\tau-2wq_1)+x_1wq_1^2\,(1-r^2)\big]\,{\Delta \bar q\,}''.
\end{gather}\end{subequations}
\subsection{The one-loop term}\label{X1}

Inserting the free propagator [i.e.\ the leading term in (\ref{G_exact})]
and $m_c^{(1)}$ into (\ref{eq_of_state}), and using the definition of $Y$ in
(\ref{Y}), we have
\begin{equation}\label{X_1}
X_1\equiv w\frac{1}{N}\sum_{\mathbf p} Y(x)-2m_c^{(1)}\,q_1r=wq_1^2\,\bar X_1
\end{equation}
where the dimensionless quantity $\bar X_1$ depends only on the three dimensionless
parameters of the theory, namely $\bar g\equiv w^2K_d/\Lambda^\epsilon$
(the dimensionless
coupling constant), $\bar \Lambda\equiv \Lambda/(2wq_1)^{1/2}$, and $x_1$:
\begin{equation}\label{bar_X_1}
\bar X_1=\bar \Lambda^\epsilon\,\bar g\,\big[ F_0^{(l)}(\bar \Lambda;r)
+x_1\, F_1^{(l)}(\bar \Lambda;r)+x_1^{\frac{d}{2}-2}\,F_0^{(s)}(r)+\dots\big].
\end{equation}
In the above formula the sub-- and superscripts of the $F$ functions correspond
to the expansion of the free propagator in Sec.\ \ref{free_propagator}:
thus $(l)$ [$(s)$] refers to the large (small) mass regime, respectively. 
Introducing the notations 
\begin{equation}\label{R_L}
G_R\equiv \frac{1}{p^2}\qquad\text{and} \qquad G_L\equiv \frac{1}{p^2+1}
\end{equation}
for the (dimensionless) ``replicon" and ``longitudinal" propagators,
Eqs.\ (\ref{G_0}) and (\ref{m_c}) give $F_0^{(l)}$:
\begin{equation}\label{F_0}
 F_0^{(l)}(\bar \Lambda;r)=4\int_0^{\bar \Lambda}dp\,p^{d-1}(-r\,G_RG_L^2+
r^3\,G_R^2G_L^2). 
\end{equation}

For obtaining $F_1^{(l)}$, one must go beyond the near infrared approximation
for $G^{xx_1}_1$, and also SBA for $Y$. A lengthy calculation gives
\begin{multline}\label{F_1}
F_1^{(l)}(\bar \Lambda;r)=4\int_0^{\bar \Lambda}dp\,p^{d-1}
\Big[
\Big(\frac{1}{2}r+\frac{1}{6}r^3\Big)\,G_RG_L^2
+\Big(-\frac{4}{15}r+\frac{7}{6}r^3\Big)
\,G_R^2G_L^2\\[4pt]
+\Big(-\frac{1}{15}r-\frac{1}{3}r^3\Big)\,G_RG_L^3+\Big(-\frac{23}{120}r
+\frac{13}{12}r^3
-\frac{57}{40}r^5\Big)\,G_R^3G_L^2+\frac{4}{15}r^6\,G_R^3G_L^3
\Big].
\end{multline}
Neglecting contributions which are irrelevant close to the critical temperature,
one can write:
\[
F_1^{(l)}(\bar \Lambda;r)=-\frac{2}{3}\,(3+r^2)\,r\,\frac{1}{\epsilon}
\,\bar \Lambda^{-\epsilon}+\bar \Lambda\text{-independent constant.}
\]
The constant is singular in six dimensions, namely it can be written as
\[
\frac{2}{3}\,(3+r^2)\,r\,\frac{1}{\epsilon}+
\Big(\frac{23}{30}\,r-\frac{13}{3}r^3+\frac{57}{10}r^5-\frac{16}{15}r^6
\Big)\,\frac{1}{\epsilon}+O(1).
\]
Unfortunately, the small mass term $F_0^{(s)}$ is too complicated to get it
in closed form. Nevertheless, its singular behavior at $d=6$ can be extracted,
and one gets a well-defined limit of $F_1^{(l)}+x_1^{-\epsilon/2}F_0^{(s)}$
in six dimensions:
\begin{equation}\label{d=6}
F_1^{(l)}(\bar \Lambda;r)+x_1^{-\epsilon/2}F_0^{(s)}(r)=
\frac{2}{3}\,(3+r^2)\,r\,\ln \bar \Lambda+
\frac{1}{2}\Big(\frac{23}{30}\,r-\frac{13}{3}r^3+\frac{57}{10}r^5
-\frac{16}{15}r^6
\Big)\,\ln x_1+O(1),\qquad d=6.
\end{equation}
\subsection{The one-loop results for the equation of state}

The first and second derivatives (with respect to $r$) of 
$F_0^{(l)}$ and the leading parts of Eqs.\ (\ref{first}) and (\ref{second}),
i.e.\ neglecting $\Delta \bar q$,
give the equations between $\tau$ and $q_1$ on the one hand:
\begin{subequations}
\begin{equation}\label{tau_vs_q1}
q_1(2\tau-2wq_1)=4\,wq_1^2\,\,\bar \Lambda^\epsilon\,\bar g\,
\int_0^{\bar \Lambda}dp\,p^{d-1}\big[G_RG_L^2-3G_R^2G_L^2\big],
\qquad\quad 6<d<8,
\end{equation}
and between
the dimensionless quantities on the other hand, in leading order:
\begin{equation}\label{x1_leading}
x_1=12\,\bar \Lambda^{\epsilon}\bar g\,\int_0^{\infty}dp\,p^{d-1}G_R^2G_L^2=
6\,\Gamma(\frac{d}{2}-2)\Gamma(4-\frac{d}{2})\,\bar \Lambda^{\epsilon}\bar g
\equiv C_d\,\bar \Lambda^{\epsilon}\bar g,
\qquad\quad 6<d<8.
\end{equation}
\end{subequations}
The diagonal mass counterterm $\delta M$ in (\ref{M_1}) can now be computed
by the help of Eqs.\ (\ref{m_c}), (\ref{tau_vs_q1}), and (\ref{x1_leading}):
\begin{equation}\label{deltaM}
\delta M = -4\,wq_1\,\,\bar \Lambda^\epsilon\,\bar g\,
\int_0^{\bar \Lambda}dp\,p^{d-1}\big[G_RG_L^2+G_R^2\big].
\end{equation}

\subsection{The $\Delta \bar q$ insertion term}

Inserting $-GM_1G$ for $G^{\text{exact}}$ into Eq.\ (\ref{eq_of_state}) with
the off-diagonal part of $M_1$ which is proportional to $\Delta \bar q$,
see Eqs.\ (\ref{G_exact}) and (\ref{M_1}), and applying the SBA (which is correct up
to this order) one arrives at:
\begin{multline*}
X_{\Delta \bar q}\equiv
-w\frac{1}{N}\sum_{\mathbf p}\,
\sum_{\gamma\not=\alpha,\beta}\,\sum_{\mu\not=\nu\not=\rho}
G_{\alpha\gamma,\mu\rho}(M_1)_{\mu\rho,\nu\rho}G_{\nu\rho,\beta\gamma}
-2m_c^{(1)}q_1\,\Delta \bar q(r)=\\
=wq_1^2\,\,\Delta \bar q(r)\,\,4\,\bar \Lambda^\epsilon\,\bar g\,
\int_0^{\bar \Lambda}dp\,p^{d-1}\big[
-G_RG_L^2+3r^2\,G_R^2G_L^2\big].
\end{multline*}

We can now use Eqs.\ (\ref{tau_vs_q1}) and (\ref{x1_leading}) in correction terms
containing $\Delta \bar q$, and also let $\bar \Lambda$ go to infinity whenever
possible (thus neglecting irrelevant contributions close to $T_c$). Somewhat
surprisingly, terms with $\Delta \bar q'$ and $\Delta \bar q''$ both
disappear due to exact cancellations when the $\Delta \bar q$ insertion term
is added to the zero-loop results in Eqs.\ (\ref{first}) and (\ref{second}).
The following simple results (correct up to the present second order
calculation) are obtained:
\begin{subequations}
\begin{gather}\label{first_add}
\frac{d}{dr}\,\big[2\tau\, q(x)+w\,q^2(x)+X_{\Delta \bar q}\big]\Bigg |_{r=1}=
q_1(2\tau-2wq_1),\\
\intertext{and}
\label{second_add}
\frac{d^2}{dr^2}\,\big[2\tau\, q(x)+w\,q^2(x)+X_{\Delta \bar q}\big]=
-2wq_1^2x_1[r-\Delta \bar q(r)].
\end{gather}
\end{subequations}

\subsection{The remaining two-loop term}

By the help of (\ref{G_exact}) and (\ref{M_1}), the basic two-loop contribution
to the equation of state in (\ref{eq_of_state}) can be written as
\begin{equation}\label{X_2}
X_2\equiv w\frac{1}{N}\sum_{\mathbf p}
\sum_{\gamma\not=\alpha,\beta}(G\Sigma G-\delta M\,G^2)_{\alpha\gamma,\beta\gamma}
-2m_c^{(2)}q_1\,r=-2 w\frac{1}{N}\sum_{\mathbf p}
(G\Sigma G-\delta M\,G^2)^{xx_1}_{\,\,\, 1}-2m_c^{(2)}q_1\,r
\end{equation}
where the last equation was obtained by the SBA, hereby neglecting terms
which are smaller by a factor of $x_1$. For making easier to display and
analyze the following --- rather complicated --- formulae, it is useful
to define the following special linear combinations of a generic ultrametric
matrix (such as $G$ or $\Sigma$, the former is used in the definitions below),
see also footnote \ref{parametrization}:
\begin{equation}\label{R_L_etc}
\begin{gathered}
G^{xx}_{R}\equiv G^{xx}_{11}-2G^{xx}_{1x_1}+G^{xx}_{x_1x_1}\,,\qquad
G^{xx}_{L}\equiv G^{xx}_{11}-4G^{xx}_{1x_1}+3G^{xx}_{x_1x_1}\,,\qquad
G^{xx}_{LA}\equiv 2G^{xx}_{1x_1}-3G^{xx}_{x_1x_1}\,;\\[3pt]
\delta G^{xx_1}\equiv G^{xx_1}_{\,\,\,1}-G^{xx_1}_{\,\,x_1}\,.
\end{gathered}
\end{equation}
Evaluating the matrix products by the SBA again, we get:
\begin{multline}\label{G_sigma_G}
(G\Sigma G-\delta M\,G^2)^{xx_1}_{\,\,\, 1}=\\[4pt]
\big[G^{xx}_{R}G^{xx_1}_{\,\,\,1}+
(2G^{xx}_{R}-G^{xx}_{L})\delta G^{xx_1}\big]\,(\Sigma^{xx}_{R}-\delta M)
-G^{xx}_{R}\delta G^{xx_1}\,(\Sigma^{xx}_{L}-\delta M)
-G^{x_1x_1}_{L}\delta G^{xx_1}\,\Sigma^{x_1x_1}_{\,\,\,x}\\[4pt]
+G^{xx}_{R}G^{x_1x_1}_{L}\,\Sigma^{xx_1}_{\,\,\,1}+
\big[G^{xx}_{R}(-G^{x_1x_1}_{\,\,\,x}+G^{x_1x_1}_{L}+G^{x_1x_1}_{LA})
-G^{xx}_{L}G^{x_1x_1}_{L}
-4\,(\delta G^{xx_1})^2\big]\,\delta \Sigma^{xx_1}\\[4pt]
+(-G^{x_1x_1}_{\,\,\,x}\delta G^{xx_1}+G^{x_1x_1}_{L}G^{xx_1}_{\,\,\,1}
+G^{x_1x_1}_{LA}\,\delta G^{xx_1})\,(\Sigma^{x_1x_1}_{L}-\delta M)
+G^{x_1x_1}_{L}\delta G^{xx_1}\,\Sigma^{x_1x_1}_{LA}.
\end{multline}
The one-loop self-energy, denoted here simply by $\Sigma$, is given
by the expression
\begin{equation}\label{sigma}
\big[\Sigma (\mathbf p)\big]_{\alpha\beta,\gamma\delta}=\frac{1}{2}w^2
\,\frac{1}{N}\sum_{\mathbf q}\,\sum_{\overset{\mu\not=\alpha,\beta}
{\nu\not=\gamma,\delta}}\Big[G_{\alpha\mu,\gamma\nu}(\mathbf q)
G_{\beta\mu,\delta\nu}(\mathbf p-\mathbf q)
+G_{\alpha\mu,\delta\nu}(\mathbf q)G_{\beta\mu,\gamma\nu}(\mathbf p-\mathbf q)
+\{\mathbf q  \leftrightarrow  \mathbf p-\mathbf q\}
\Big]
\end{equation}
where the third and fourth terms inside the curly brackets, i.e.\ 
$\{\mathbf q  \leftrightarrow  \mathbf p-\mathbf q\}$, are the same as the
first and second ones, but with $\mathbf q$ and $\mathbf p-\mathbf q$
transposed.

\section{Calculation of the correction to the order parameter function}
\label{results}

In this section, the basic formulae for the calculation of the order
parameter function $q(x)$ in two-loop order are presented. In what follows
$q_1$, $x_1$, and $\Delta \bar q(r)$ --- with $r=x/x_1$,
and $\Delta \bar q(1)=0$ exactly --- are expressed by the parameters of
the model in (\ref{simple_L}),
namely $\tau$ (temperature), $w$ (coupling constant),
and $\Lambda$ (momentum cutoff).

The equation between $q_1$ and $\tau$ is obtained by the first derivative
of Eq.\ (\ref{eq_of_state}) evaluating at $x=x_1$:
\begin{subequations}
\begin{equation}\label{basic_1}
\tau-wq_1=-\frac{1}{2q_1}\,\left[\frac{d}{dr}(X_1+X_2)\right]_{r=1}
\end{equation}
where Eq.\ (\ref{first_add}) and the definitions in (\ref{X_1}) and
(\ref{X_2}) were used. The second derivative of (\ref{eq_of_state})
together with Eq.\ (\ref{second_add}) provides $x_1$ as:
\begin{equation}\label{basic_2}
x_1=\frac{1}{2wq_1^2}\,\left[\frac{d^2}{dr^2}(X_1+X_2)\right]_{r=1}.
\end{equation}
Using (\ref{second_add}) again, but now for generic $r$, and also
(\ref{basic_2}), one gets the correction to the leading, purely linear
order parameter function:
\begin{equation}\label{basic_3}
\Delta \bar q(r)=\frac{1}{2wq_1^2x_1}\,
\left\{\left[\frac{d^2}{dr^2}(X_1+X_2)\right]_{r=1}\times r
-\left[\frac{d^2}{dr^2}(X_1+X_2)\right]\right\}.
\end{equation}
\end{subequations}

\subsection{The calculation of $x_1$}\label{result_x1}

\subsubsection{Generic dimensions $6<d<8$}
Applying the results of subsection \ref{X1} and Appendix \ref{C}, one can
write for $x_1$ in Eq.\ (\ref{basic_2}):
\begin{equation}\label{x_1_above}
\begin{split}
x_1&=\frac{1}{2}\,\bar\Lambda^{\epsilon}\bar g\,\frac{d^2}{dr^2}
\Big\{F_0^{(l)}+x_1\,\big[F_1^{(l)}+x_1^{-\epsilon/2}F_0^{(s)}\big]\Big\}
\Bigg |_{r=1}
+\frac{1}{2wq_1^2}\,\left[\frac{d^2}{dr^2}X_2\right]_{r=1}\\[5pt]
&=C_d\,\bar\Lambda^{\epsilon}\bar g+(K_1\,\bar\Lambda^{-\epsilon}+K_2+K_3\,
x_1^{-\epsilon/2})\,x_1^2,
\end{split}
\end{equation}
with some $d$-dependent constants $K_1$, $K_2$, and $K_3$. Since
one can use the leading behavior $\bar\Lambda^{\epsilon}\bar g
\sim x_1$ in the correction terms, these constants get contributions from both
$X_1$ and $X_2$. Rearranging this equation provides $\bar\Lambda^{\epsilon}
\bar g$ in terms of $x_1$ and $\bar g$:
\begin{equation}\label{x_1_above_v}\begin{aligned}
\bar\Lambda^{\epsilon}\bar g&=(C_d^{-1}-K_1\,\bar g+\dots)\,x_1+(-C_d^{-1}K_2
+\dots)\,x_1^2+(-C_d^{-1}\,K_3+\dots)\,x_1^{2-\epsilon/2}\\[4pt]
&\equiv f^{(1)}(\bar g)\,x_1+f^{(2)}(\bar g)\,x_1^2+f^{(na)}(\bar g)\,
x_1^{2-\epsilon/2}\equiv \mathcal F(x_1,\bar g).
\end{aligned}
\end{equation}
The left-hand side is a measure of the distance from the critical temperature:
$\bar\Lambda^{\epsilon}\bar g\sim \tau^{-\epsilon/2}$, note that $\epsilon<0$,
and the temperature dependence of $x_1$ is obtained by finding the root of the
above equation. It can be easily seen that,
when approaching the critical temperature, the ratio
of the correction term of
$x_1$ to the leading one is proportional to $\bar g$.
Although this is the usual behavior above the upper critical
dimension, the nonanalytical term, i.e.\ $f^{(na)}(\bar g)\,x_1^{2-\epsilon/2}$,
is a peculiarity of the spin glass field theory. 

\subsubsection{At the upper critical dimension: $d=6$}
One can evaluate (\ref{x_1_above}) at exactly $\epsilon=0$ by using
Eq.\ (\ref{d=6}) for the $X_1$ part, furthermore Eqs.\  (\ref{X_2^1}), (\ref{X_2^2}),
(\ref{X_2^3}),
(\ref{X_2^4}), (\ref{X_2^5}), and
(\ref{X_2^6}) from Appendix \ref{C} for the $X_2$ part: 
\begin{equation}\label{x_1_at}
x_1=6\,\bar g -\frac{1}{3}x_1^2\big[\ln \bar\Lambda+2\ln x_1+O(1)\big]
=6\,\bar g-12\,\bar g^2\big[\ln \bar\Lambda+2\ln \bar g+O(1)\big]+\dots
\end{equation}
where $C_{d=6}=6$ has been used.%
\footnote{The $x_1^2\,\ln x_1$ contributions from $(d^2/dr^2\,X_1)_{r=1}$ and
$(d^2/dr^2\,X_2)_{r=1}$ with the $\Sigma_R^{xx}$ self-energy insertion cancel
each other --- a surprising effect whose origin is not understood. The
$-2/3\,x_1^2\,\ln x_1$ above comes from the $\Sigma_L^{xx}$,
$\Sigma^{x_1x_1}_{\,\,\,x}$, and $\Sigma^{xx_1}_{\,\,\,1}$ self-energy components.}
Evidently, the correction term blows up
when approaching the critical temperature, demonstrating the well-known phenomenon
that the simple perturbative method breaks down at the upper critical dimension.
Nevertheless, one can make a comparison
with the expanded form of the renormalization group
result in Eq.\ (34) of Ref.\ \cite{PT}:
\[
x_1=6\,\left[\frac{\bar g}{1-\bar g\ln (\tau \bar g^{-5/3}/\Lambda^2)}\right]+\dots
=6\,\bar g-12\,\bar g^2\ln \bar\Lambda-10\, \bar g^2\ln \bar g+\dots
\]
where the equality, in leading order,
of $\ln \bar\Lambda\equiv \ln\big(\Lambda/\sqrt{2wq_1}\big)$
with $-\ln(2\tau/\Lambda^2)/2$ has been used.
The discrepancy in the $\bar g^2\ln \bar g$
term arises from two factors: Firstly, from the fact that only the linear leading term
of the scaling function $\hat x_1(x)$ was computed in \cite{PT} which is,
however, sufficient for getting the $\bar g^2\,\ln \tau$ contribution. Secondly,
the $\bar g^2\ln \bar g$ term in (\ref{x_1_at}) comes from the small mass regime,
and it is not expected to be controlled by renormalization around the critical
theory.
Considering that in Ref.\ \cite{PT} $x_1$ was computed by an expansion around
the replica symmetric field theory --- in contrast to the present fully
RSB calculation ---, the agreement found in the $\bar g^2\,\ln \tau$
term is rather reassuring.

\subsubsection{Below the upper critical dimension: $d=6-\epsilon$}
One can revive perturbation theory below six dimensions by the method
of the $\epsilon$-expansion: the $d$-dimensional integrals are expanded
in $\epsilon$, and the coupling constant $\bar g$ must be specially chosen
for having correctly exponentiating logarithms.
In this second order calculation, one can
stay in six dimensions when evaluating the correction terms. As a result,
Eq.\ (\ref{x_1_at}) is only slightly modified:
\begin{equation}\label{x_1_below}
x_1=6\,\bar g\,(1+\epsilon\ln \bar\Lambda)
-\frac{1}{3}x_1^2\big[\ln \bar\Lambda+2\ln x_1+O(1)\big]
=6\,\bar g+6\,\bar g\,(\epsilon-2\,\bar g)\,\ln \bar\Lambda
-\frac{1}{3}x_1^2\big[2\ln x_1+O(1)\big]
\end{equation}
where $C_d=6+O(\epsilon^2)$ was used. The correction term becomes infinite
at criticality, except for $2\,\bar g=\epsilon+O(\epsilon^2)$ which is just
the fixed point condition in the first order of the perturbative renormalization
group, see for instance \cite{PT}.\footnote{\label{g_hat}%
More importantly, we will see in the next subsection that the very same condition
ensures that the $\ln \bar\Lambda$'s correctly exponentiate, at least in the order
we are studying here, 
when the $\tau$ versus $q_1$ relation is computed perturbatively
below six dimensions. One may expect that a choice $\bar g=\hat{\bar g}$ will
guarantee in {\em any order\/}: (i) the vanishing of the $\ln \bar\Lambda$'s
in the series for $x_1$, and (ii) the correct exponentiation of them in the
$\tau$ versus $q_1$ series. Nevertheless, $\hat{\bar g}$ is not needed to be
identical with the fixed point, except in first order in $\epsilon$,
since $\bar g$ is a {\em bare\/} dimensionless coupling.}
In this way, the equation
that determines $x_1$ becomes independent of the temperature, and $x_1$ itself
is nonzero and universal below six dimensions:
\begin{equation}\label{x_1_below_2}
3\,\epsilon=x_1+\frac{1}{3}x_1^2\big[2\ln x_1+O(1)\big].
\end{equation}
In a generic dimension $d<6$, there should exist a function $\hat{\mathcal
F}(x_1,\bar g)$ such that $x_1$ is determined from
\begin{equation}\label{x_1_below_3}
\hat{\mathcal F}(x_1,\hat{\bar g})=0
\end{equation}
where $\hat{\bar g}$ is the special coupling as explained in footnote
\ref{g_hat} .

\subsection{The Edwards-Anderson order parameter}\label{result_q1}

\subsubsection{$q_1$ in dimensions $6<d<8$}

After evaluating Eq.\ (\ref{basic_1}), one gets:
\begin{equation}\label{q_1_above}
\begin{split}
&\tau-wq_1=\\[1pt]
&=-\frac{1}{2}(wq_1)\,\bar\Lambda^{\epsilon}\bar g\,
\Big\{-4\,\int_0^{\bar \Lambda}dp\,p^{d-1}G_RG_L^2+C_d
+x_1\,\frac{d}{dr}
\big[F_1^{(l)}+x_1^{-\epsilon/2}F_0^{(s)}\big]_{r=1}\Big\}
-\frac{1}{2q_1}\,\left[\frac{d}{dr}X_2\right]_{r=1}\\[7pt]
&=(wq_1)\,\Big\{\left(-\frac{2}{\epsilon}+
\frac{1-2\epsilon}{3\epsilon}\,C_d\,\bar\Lambda^{\epsilon}\right)\,\bar g
+\big(K_1'\,\bar\Lambda^{-2\epsilon}+K_2'\,\bar\Lambda^{-\epsilon}+K_3'+K_4'\,
x_1^{-\epsilon/2}\big)\,x_1^2
\Big\}\\[7pt]
&=(wq_1)\,\Big\{\left(-\frac{2}{\epsilon}+
\frac{1-2\epsilon}{3\epsilon}\,C_d\,\bar\Lambda^{\epsilon}\right)\,\bar g
+\big(K_1'+K_2'\,\bar\Lambda^{\epsilon}+K_3'\,\bar\Lambda^{2\epsilon})
\,C_d^2\,\,\bar g^2+K_4'\,x_1^{2-\epsilon/2}\Big\},
\end{split}
\end{equation}
with some $d$-dependent constants $K_1'$,\dots $K_4'$ [see the remark
below Eq.\ (\ref{x_1_above}) which is appropriate here too]. The terms first
and second order in
the dimensionless coupling constant $\bar g$ in the last line are quite usual
in the perturbative expansion of an order parameter above the upper critical
dimension, whereas the nonanalytic last term has its origin in the far infrared regime,
and is specific to the spin glass field theory with the two distinct bands
of the mass spectrum.

\subsubsection{$q_1$ in six dimensions}

For finding $q_1$ {\em perturbatively\/} in $d=6$, one can use (\ref{d=6}):
\[
\frac{d}{dr}
\big[F_1^{(l)}+x_1^{-\epsilon/2}F_0^{(s)}\big]_{r=1}=4\ln \bar\Lambda
+\frac{74}{15}\ln x_1+O(1),
\]
and collect terms from 
(\ref{X_2^1}), (\ref{X_2^2}),
(\ref{X_2^3}),
(\ref{X_2^4}), (\ref{X_2^5}), and
(\ref{X_2^6}):
\[
\frac{1}{q_1}\,\left[\frac{d}{dr}X_2\right]_{r=1}=(wq_1)\,x_1^2\,
\Big[-\frac{2}{3}\,\ln x_1+O(1)\Big].
\]
[It is remarkable that the different terms for $\ln^2 \bar\Lambda$ and 
$\ln \bar\Lambda$ in $\big(\frac{d}{dr}\,X_2\big)_{r=1}$ all cancel each other, and the
$-2/3\,\ln x_1$ comes from the $\Sigma^{xx}_R$ component, all the other
contributions are zero.]
Putting these second order terms into (\ref{q_1_above}), and evaluating the
first order one at $d=6$, one gets:
\[
\tau-wq_1=(wq_1)\,\Big\{(2\ln \bar\Lambda-4)\,\bar g-6\Big[2\ln \bar\Lambda+
\frac{7}{15}\ln \bar g+O(1)\Big]\,\bar g^2\Big\}
\]
where $\ln x_1$ was substituted by $\ln(6\bar g)$. This equation gives $\tau$
versus $q_1$ for a fixed temperature below $T_c$. For getting the behavior
of $q_1$ for $\tau$ approaching zero for a given system (i.e.\ for some given $\bar g$),
one must resum an infinite series: this is just done by the renormalization group,
as in Ref.\ \cite{PT}.

\subsubsection{$\epsilon$-expansion for $q_1$, $d<6$}\label{exp1}

The first term in the second row of Eq.\ (\ref{q_1_above})
must now be expanded up to
$O(\epsilon)$, while the second and third ones can be evaluated at
six dimensions. We get for $\tau$:
\begin{equation}\label{q_1_below}
\begin{split}
\tau&=
(wq_1)\,\Big\{
1+2\,(\ln \bar\Lambda-2)\,\bar g+\Big(\ln^2\bar\Lambda -4\ln\bar\Lambda+
\frac{\pi^2}{12}\Big)\,\epsilon \bar g-6\Big[2\ln \bar\Lambda+\frac{7}{15}
\ln \bar g+O(1)\Big]\,\bar g^2
\Big\}\\[5pt]
&=(wq_1)\,\Big\{(1-4\bar g)+(2\bar g-4\epsilon \bar g-12\bar g^2)\,\ln \bar\Lambda
+\epsilon \bar g\,\ln^2\bar\Lambda
+\Big[-\frac{14}{5}\,\bar g^2\,\ln \bar g
+\frac{\pi^2}{12}\,\epsilon \bar g+O(\bar g^2)\Big]
\Big\}.
\end{split}
\end{equation}
The first three terms in the curly brackets can be considered as part of
the expansion of
$(1-4\bar g+\dots)\,\bar\Lambda^\kappa$, provided that
$(1-4\bar g)\kappa=2\bar g-4\epsilon \bar g-12\bar g^2$
and $\frac{1}{2}\kappa^2=\epsilon \bar g$, yielding the exponentiation condition
$\bar g=\hat{\bar g}=\frac{1}{2}\epsilon$ --- which coincides with
the requirement of vanishing $\ln \bar\Lambda$ contribution to $x_1$
found in the previous subsection, see footnote \ref{g_hat} ---, and
$\kappa=2\hat{\bar g}-12\hat{\bar g}^2+O(\hat{\bar g}^3)$.
Identifying the critical exponent $\beta$ as $\beta^{-1}=1-\frac{1}{2}\kappa$,
it is obtained:
\begin{equation}\label{beta}
\beta=1+\hat{\bar g}-5\hat{\bar g}^2+\dots=
1+\frac{1}{2}\epsilon+O(\epsilon^2).
\end{equation}
Unfortunately only the leading behavior of $\hat{\bar g}$ is available
at the moment, preventing us from computing $\beta$ up to $O(\epsilon^2)$,
and compare it with known results from renormalization in the
symmetric (high-temperature) theory \cite{Alcantara_Bonfim,Gr85}. (See
also Ref.\ \cite{beyond} where this first order result for $\beta$ has been presented
from calculation in the glassy phase.)

\subsection{The correction to the order parameter function: $\Delta\bar q(r)$}
\label{result_delta_q}

Eq.\ (\ref{basic_3}) shows that $\Delta\bar q(1)=0$, as it must be. Furthermore,
one can conclude from this formula that terms linear ($\sim r$) and cubic
($\sim r^3$) in $X_1$ and $X_2$ give no contributions to $\Delta\bar q(r)$.
Using Eqs.\ (\ref{X_1}), (\ref{bar_X_1}), (\ref{F_0}), and (\ref{F_1}),
it then immediately follows:
\begin{equation}\label{delta_q}
\begin{split}
\Delta \bar q(r)&=\frac{1}{2}\,\bar\Lambda^{\epsilon}\bar g
\left\{\left[\frac{d^2}{dr^2}\big(F_1^{(l)}+x_1^{-\epsilon/2}F_0^{(s)}\big)
\right]_{r=1}\times r
-\left[\frac{d^2}{dr^2}\big(F_1^{(l)}+x_1^{-\epsilon/2}F_0^{(s)}\big)
\right]\right\}
\\[5pt]
&+\frac{1}{2wq_1^2x_1}\,
\left\{\left[\frac{d^2}{dr^2}X_2\right]_{r=1}\times r
-\left[\frac{d^2}{dr^2}X_2\right]\right\}.
\end{split}
\end{equation}

\begin{itemize}
\item 
As far as a generic dimension $6<d<8$ is concerned, one can compute the
first part of (\ref{delta_q}) by use of Eq.\ (\ref{F_1}):
\[
\Delta \bar q(r)=x_1\,C_d^{-1}\Big[-57(r-r^3)\,\int_0^\infty dp\,p^{d-1}
G_R^3G_L^2+16(r-r^4)\,\int_0^\infty dp\,p^{d-1}G_R^3G_L^3\Big]
+O\big(x_1^{1-\frac{\epsilon}{2}}\big).
\]
Only terms proportional to $r^5$ give contributions to the second part
of (\ref{delta_q}) --- these come from $\Sigma^{xx}_{R}$,
$\Sigma^{xx}_{L}$, $\Sigma^{x_1x_1}_{\,\,\,x}$, $\Sigma^{xx_1}_{\,\,\,1}$,
and $\delta \Sigma^{xx_1}$ ---, yielding
\[
\frac{1}{2wq_1^2x_1}\,
\left\{\left[\frac{d^2}{dr^2}X_2\right]_{r=1}\times r
-\left[\frac{d^2}{dr^2}X_2\right]\right\}
\sim x_1\,(r-r^3)\,\times [\text{convergent 2-loop integrals}],
\]
and also a complicated nonanalytical contribution of order
$x_1^{1-\frac{\epsilon}{2}}$, coming from integrals in the far infrared
region, which is negligible for $d>6$, but becomes more and more important
when approaching $d=6$.

One can conclude from this 2-loop calculation that the order parameter
function takes the form
\[
q(x)/q_1=(1+a_d\,x_1+\dots)\,r+x_1\,(c_d\,r^3+d_d\,r^4)+
\Delta \bar q^{(na)}(r)+\dots,
\quad r=x/x_1,\quad 6<d<8
\]
where the nonanalytical contribution is subleading
in this dimensional regime:
\[
\Delta \bar q^{(na)}(r)\sim x_1^{1-\frac{\epsilon}{2}}.
\]
All the temperature dependence of $q(x)$ is absorbed into $q_1$ and $x_1$
which are, as it follows from our scheme, the {\em exact\/} Edwards-Anderson
order parameter and breakpoint of $q(x)$. Furthermore, although $q(x)/q_1$
should, in principle, depend on both $x_1$ and $\bar g$, it proved to be,
at least up to the order considered, $\bar g$-independent.

The emergence of the $x_1\times d_d\,r^4$ term, coming solely from the one-loop
graph $X_1$, may seem somewhat surprising, although a similar $x^4$
contribution has been found in the mean field order parameter function
\cite{Temesvari89}. (But see Sec.\ \ref{conclusion} for confronting
the results here for $6<d<8$ with their mean field counterparts,
i.e.\ $d=\infty$.)

\item
In exactly six dimensions, (\ref{delta_q}) can be evaluated by use of
Eqs.\ (\ref{d=6}), (\ref{X_2^1}), (\ref{X_2^2}), (\ref{X_2^3}), and
(\ref{X_2^4}); note that only terms proportional to $r^5$
and the sole $r^6$ contribution from $X_1$ give nonvanishing result:
\[
\Delta \bar q(r)=-\Big[\frac{1}{4}(r-r^3)+\frac{4}{3}(r-r^4)
\Big]\,x_1\ln x_1+O(x_1),\qquad\quad d=6.
\]
\end{itemize}

\section{The study of the momentum-dependent mass}\label{mass}

By Dyson's equation, the inverse of the exact propagator is identical with
the mass operator $\Gamma(\mathbf p)$, and stability of the RSB phase
demands that the eigenvalues of
$\Gamma(\mathbf p=\mathbf 0)$ be all nonnegative.
Eqs.\ (\ref{Gamma}) and (\ref{M_1}) give the elements of the mass operator as:
\begin{equation}\label{Gamma_generic}
\begin{aligned}
\Gamma_{\alpha\beta,\alpha\beta}&=p^2-2wq_1+x_1wq_1+\delta M
-\Sigma_{\alpha\beta,\alpha\beta}(\mathbf p),\\[4pt]
\Gamma_{\alpha\gamma,\beta\gamma}&=-wq_{\alpha\beta}-
\Sigma_{\alpha\gamma,\beta\gamma}(\mathbf p),\\[4pt]
\Gamma_{\alpha\beta,\gamma\delta}&=
-\Sigma_{\alpha\beta,\gamma\delta}(\mathbf p),
\end{aligned}
\end{equation}
and everything above is understood to be exact quantity (the self-energy,
for instance, although the same notation is used as for its first order
part throughout the paper). Instead of a full analysis, our study will be
confined to the highest longitudinal eigenvalue and to the family of the
replicon ones.

\subsection{The longitudinal mass}\label{exp2}

Applying the results from Ref.\ \cite{block_diag}, an eigenvector 
$f_{\alpha\beta}$ of the longitudinal subspace has the same ultrametric
structure
as the order parameter $q_{\alpha\beta}$, and we can use a similar 
parametrization for it, i.e.
\[
f_{\alpha\beta}=f_{\alpha\cap\beta}=f(x)=f_1\,\bar f(r)=
f_1\,[r+\Delta \bar f(r)],
\quad\text{where}\quad r=x/x_1, \quad\text{and}\quad
\Delta \bar f(1)=0.
\]
The eigenvalue equation can now be written as
\begin{multline*}
\frac{1}{2}\sum_{\gamma\not=\delta}\Gamma_{\alpha\beta,\gamma\delta}
\,f_{\gamma\delta}=(p^2-2wq_1+x_1wq_1+\delta M)\,\bar f(r)
+2wq_1\,\big[\bar q(r)+\bar f(r)-x_1\{\bar q(r),\bar f(r)\}\big]\\[5pt]
-\Sigma^{xx}_R(\mathbf p)\,\bar f(r)+2\delta \Sigma^{xx_1}(\mathbf p)=
\lambda \,\bar f(r).
\end{multline*}
The SBA has been used in the terms with the one-loop self-energy components,
which is correct at the present level of approximation. It is easy to
check that $\lambda=p^2+2wq_1$ and $\bar f(r)=r$ are the zeroth order
solutions. Due to the $x_1$ in front of the bilinear form, see (\ref{bilinear})
for its definition, it is sufficient to replace  $\bar q$ and $\bar f$ by $r$:
$\{\bar q(r),\bar f(r)\}\approx \{r,r\}=r-r^3/3$.
Setting $r=1$ and using Eq.\ (\ref{Sigma1_p0}) yield the first order result
for the longitudinal momentum-dependent mass:
\begin{equation}\label{Gamma_long}
\lambda\equiv \Gamma_{\text{long}}(\mathbf p)=
p^2+2wq_1-\frac{4}{3}x_1wq_1-\big[\Sigma^{x_1x_1}_R(\mathbf p)
-\Sigma^{x_1x_1}_R(\mathbf p=\mathbf 0)\big]+2\delta \Sigma^{x_1x_1}(\mathbf p).
\end{equation}
By the help of Eqs.\ (\ref{sigma_generic}) and (\ref{sigma_na}), and the row
for $\delta \Sigma^{xx_1}$ in Table \ref{table_sigma}, the (zero-momentum)
longitudinal mass above six dimensions is as follows:
\begin{equation}\label{Gamma_long_above}
\Gamma_{\text{long}}(\mathbf p=\mathbf 0)=2wq_1\,\Big[
1+\frac{2}{\epsilon}\,\bar g+O(x_1)+O(x_1^{1-\epsilon/2})\Big],
\qquad 6<d<8.
\end{equation}

Considering that the longitudinal momentum-dependent mass is the inverse
of the longitudinal exact propagator, one can expect that its behavior
below six dimensions is
governed by the critical fixed point, and it has the following scaling form:
\begin{equation}\label{Gamma_scaling}
\Gamma_{\text{long}}(\mathbf p)=p^2\,(p/\Lambda)^{-\eta}\,
\mathcal G\big[(p/\Lambda)^2\bar\Lambda^{4\nu/\beta}\big],\qquad d<6.
\end{equation}
One can check this scaling by evaluating the self-energy
components in (\ref{Gamma_long}) at exactly six dimensions:
\begin{multline*}
\Sigma^{xx}_R(\mathbf p)-\Sigma^{xx}_R(\mathbf p=\mathbf 0)=
\bar g\,wq_1\,\Big\{
\frac{1}{9}(p^2/2wq_1)\big[12\ln (\Lambda/p)+11\big]
-\big[4\ln (\Lambda/p)+1\big]+4(\ln \bar\Lambda-2)+6(1-r^2)\\[5pt]
+2(1+2r^2)\,(p^2/2wq_1)^{-1}\ln(p^2/2wq_1)+(3+4r^2)\,(p^2/2wq_1)^{-1}+
O\big[(p^2/2wq_1)^{-2}\big]
\Big\},
\end{multline*}
and
\begin{multline*}
\delta \Sigma^{x_1x_1}(\mathbf p)=\\[5pt]
\bar g\,wq_1\,\Big\{
-\frac{1}{2}\big[4\ln (\Lambda/p)+1\big]-5\,(p^2/2wq_1)^{-1}\ln(p^2/2wq_1)
-\frac{1}{2}(p^2/2wq_1)^{-1}+O\big[(p^2/2wq_1)^{-2}\big]
\Big\}.
\end{multline*}
Inserting these expressions into (\ref{Gamma_long}) and using $x_1=6\bar g$,
one gets:
\[
\eta=-\frac{2}{3}\bar g,\quad \frac{4\nu}{\beta}=2+\frac{4}{3}\bar g,\quad
\text{and}\quad \mathcal G(u)=(1+u^{-1})+\Big[-\frac{11}{18}
-\frac{1}{3}u^{-1}\ln u-4u^{-2}(2\ln u+1)+\dots\Big]\,\bar g,
\]
in full agreement with the $\epsilon$-expansion results from calculations
in the symmetric (high temperature) phase in Ref.\ \cite{Gr85}, whenever
$\bar g$ is substituted by $\hat{\bar g}=\epsilon/2$, namely
\[ 
\eta=-\frac{1}{3}\epsilon,\quad \nu=\frac{1}{2}(1+\frac{5}{6}\epsilon),
\quad\text{and}\quad
\beta=1+\frac{1}{2}\epsilon.
\]

The zero-momentum limit of $\Gamma_{\text{long}}$ is the inverse longitudinal
susceptibility. By use of Eq.\ (\ref{Gamma_scaling}), one gets its behavior:
\[
\Gamma_{\text{long}}(\mathbf p=\mathbf 0)\sim 
\bar\Lambda^{\frac{2\nu(\eta-2)}{\beta}}\sim \bar\Lambda^{\frac{-2\gamma}{\beta}}
\sim \tau^{\gamma},\qquad d<6.
\]
By means of Eqs.\ (\ref{Sigma5}) and (\ref{propagator_list}), it follows
that
\[
\delta \Sigma^{x_1x_1}(\mathbf 0)=2C_d^{-1}x_1(wq_1)
\int_0^{\bar \Lambda}dp\,p^{d-1}\big(-G_R^2G_L-8G_R^2G_L^2+8G_R^2G_L^3\big)
\overset{d=6}{=}-\frac{1}{3}\,x_1(wq_1)\,(\ln \bar \Lambda+2)
\]
where the final result was obtained by neglecting irrelevant, i.e.\ 
$\sim \bar \Lambda^{-2}$, terms. The zero-momentum limit of (\ref{Gamma_long})
can now be written as:
\[
2wq_1-\frac{8}{3}x_1(wq_1)-\frac{2}{3}x_1(wq_1)\,\ln \bar \Lambda
=2wq_1\,(1-8\bar g)\bar \Lambda^{-2\bar g},\qquad x_1=6\bar g+\dots\quad.
\]
One can then conclude that $\gamma/\beta=1+\bar g$ and, using (\ref{beta}),
$\gamma=1+2\bar g$. This yields --- again in full agreement with
Ref.\ \cite{Gr85} --- $\gamma=1+\epsilon$, after the special condition
$\bar g=\hat{\bar g}$ for the coupling constant has been applied.

\subsection{The replicon band of $\Gamma(\mathbf  p)$}\label{replicon}

It was shown in Refs.\ \cite{block_diag,beyond} that the replicon eigenvalues
of any ultrametric matrix can be easily computed by direct substitution
of the matrix elements into an expression such as Eq.\ (41)
in \cite{block_diag}. Inserting the components of $\Gamma(\mathbf  p)$
in Eq.\ (\ref{Gamma_generic}) into this formula, and keeping terms up to
first order in $x_1$, a surprisingly simple result is obtained:
\[
\Gamma_{\text{repl}}(x;u,v)=p^2+x_1\,(wq_1)\,\big[
(r_1^2+r_2^2)/2-r^2\big]-
\big[\Sigma^{xx}_R(\mathbf p)-\Sigma^{xx}_R(\mathbf p=\mathbf 0)\big]
\]
where $0\le r=x/x_1 \le r_1=u/x_1,\, r_2=v/x_1 \le 1$, and Eq.\ 
(\ref{Sigma1_p0}) was applied. The middle term is just the zero-momentum
replicon mass; marginal stability is clearly demonstrated. The $u=v=x$ mode
is known exactly \cite{reparametrization} being
a zero-(Goldstone)mode, and we can see
this here perturbatively:
\[
\Gamma_{\text{repl}}(x;x,x)=0+O(x_1^2),\qquad 0\le x \le x_1,
\qquad \mathbf p=\mathbf 0.
\]
One can suspect that this marginality persists, and is satisfied
order by order in the perturbation expansion.

\section{Discussion of the results, and some conclusions}\label{conclusion}

The glassy phase of the replica field theory representing the Ising spin glass
has a special dimensionless parameter, $x_1$, not present in ordinary field
theories, which is a characteristic of the RSB low-temperature phase. Just
below $T_c$ and in systems above the upper critical dimension (more precisely
for $6<d<8$) $x_1$ is related to the other two dimensionless quantities, namely
$\bar \Lambda=\Lambda/(2wq_1)^{\frac{1}{2}}$ (which diverges at criticality)
and $\bar g=w^2K_d/\Lambda^{\epsilon}$
(the dimensionless coupling constant), by an equation like (\ref{x_1_above})
or (\ref{x_1_above_v}). It has been shown in Ref.\ \cite{PT} that $x_1$
becomes nonzero at criticality, i.e.\ it is independent of $\bar \Lambda$,
and universal below the upper critical dimension.
The $d$-dependence of this universal value is calculated in this
paper up to second order in $\epsilon$, see (\ref{x_1_below_2}). The equation
for generic $d$ can be written as in Eq.\ (\ref{x_1_below_3}), where the special
coupling constant $\hat{\bar g}$ insures proper exponentiation of temperature
singularities, and it is related, but not equivalent, to the fixed point of the
Wilson-type perturbative renormalization group.

In the classical perturbative regime above the upper critical temperature
the behavior of $\tau/(wq_1)$ and $\Gamma_{\text{long}}(\mathbf p=\mathbf 0)
/(2wq_1)$ are displayed in Eqs.\ (\ref{q_1_above}) and (\ref{Gamma_long_above}).
They both are the sum of a regular and an anomalous term. The regular ones have
the following common structure: $\sum_{i,j}c_{ij}\bar g^ix_1^j$, and the $c_{ij}$
coefficients for $i+j=L$ can be calculated by an $L$-loop calculation (higher
loop terms do not change the result), notwithstanding that the free propagator
itself is an infinite series in $x_1$. On the other hand, the anomalous term
is nonanalytic in $x_1$: it is proportional to $x_1^{2-\epsilon/2}$ and
$x_1^{1-\epsilon/2}$ in the two cases. It is argued in the following that this
anomalous contribution (which comes always from far infrared integration) is
nonperturbative in the sense that higher loop graphs yield similar
anomalous terms. Let us look at, as an example,
the following $k+1$-loop contribution to the equation of state in
(\ref{eq_of_state}), see also (\ref{G_exact}):
\begin{align*}
X^{(k)}&\equiv
w\frac{1}{N}\sum_{\mathbf p}
\sum_{\gamma\not=\alpha,\beta}\big\{G[(\Sigma-M_1)G]^k\big\}
_{\alpha\gamma,\beta\gamma}\\[5pt]
&=-2 w\frac{1}{N}\sum_{\mathbf p}
\big[G^{xx}_{R}G^{xx_1}_{\,\,\,1}+
(2G^{xx}_{R}-G^{xx}_{L})\delta G^{xx_1}\big]\,\big\{[(\Sigma-M_1)G]^kG^{-1}
\big\}^{xx}_R
\end{align*}
where the SBA was used, and only the replicon contribution is considered here,
see also (\ref{G_sigma_G}).
Since $\{\dots\}^{xx}_R$ is an exact eigenvalue of {\em any generic\/} ultrametric
matrix, we can write:
\[
\big\{[(\Sigma-M_1)G]^kG^{-1}\big\}^{xx}_R=\big[(\Sigma-M_1)^{xx}_R\big]^k
\times \big[G^{xx}_R\big]^{k-1}.
\]
By the help of Eqs.\ (\ref{M_1}), (\ref{Sigma1_p0}), (\ref{sigma_generic}),
(\ref{sigma_na}), and Table \ref{table_sigma}, one gets for momenta in the
far infrared, i.e.\  $\frac{p^2}{x_1(2wq_1)}=u$ with $u=O(1)$:
\begin{align*}
(\Sigma-M_1)^{xx}_R&=x_1\,(wq_1)\big\{r^2+
[f(ux_1)-f(0)]\,{\bar\Lambda}^{-\epsilon}+[\sigma^{(a)}(ux_1)-\sigma^{(a)}(0)]+
x_1^{1-\epsilon/2}\,[f^{(s)}(u)-f(0)]
\big\}\\[4pt]
&\approx x_1\,(wq_1)\,r^2,
\end{align*}
the approximation meaning leading order in $x_1$. For the replicon free propagator
one has exactly:
\[
G^{xx}_R=\frac{1}{p^2+\lambda_0(x;x_1,x_1)}=\frac{1}{p^2+x_1\,wq_1}
\]
where Eq.\ (\ref{free_replicon}) has been used for getting the last equality.
Finally, using the classification of far infrared propagators at the end of
Appendix \ref{A} and Eq.\ (\ref{G_small_mass}), $X^{(k)}$ has the following
leading contribution in the small mass regime:
\begin{align*}
X^{(k)}&=-2 w\frac{1}{N}\sum_{\mathbf p}\frac{1}{x_1\,p^4}\,G^{(s)}\!\!
\left[\frac{p^2}{x_1\,(2wq_1)}\right](x_1wq_1)^k\,r^{2k}
\,\frac{1}{(p^2+x_1\,wq_1)^{k-1}}\\[4pt]
&\sim (wq_1^2)\,x_1^{1-\epsilon/2}\,(\bar g{\bar\Lambda}^{\epsilon})\,r^{2k}\sim 
(wq_1^2)\,x_1^{2-\epsilon/2}\,r^{2k}
\end{align*}
where the last formula was obtained by use of (\ref{x1_leading}). The order of
this formula, i.e.\ the power of $x_1$, is
independent of the number of loops $k+1$, and $\tau/(wq_1)$ gets a nonanalytic
contribution which is proportional to $x_1^{2-\epsilon/2}$ in any order of
the loop-expansion.

What has been learnt from the study of the classical perturbative regime is
extensible to the case of $d<6$. 
$\tau/(wq_1)$ and $\Gamma_{\text{long}}(\mathbf p=\mathbf 0)$ can be separated
into a regular part (coming from integration in the near infrared), and an
anomalous one originating from the small mass (far infrared) sector. The regular
part is under the control of the critical fixed point, and usual critical
exponents can be computed, after the separation has been done, as 
in \ref{exp1} and \ref{exp2}. The dangerous infrared behavior is cut off
by the small mass, and it is isolated into the anomalous part; see as an example
the $\bar g^2\,\ln \bar g$ term in Eq.\ (\ref{q_1_below}).

$\Delta \bar q(r)=q(x)/q_1-r$, $r=x/x_1$, has been assumed in Ref.\ \cite{PT}
to be proportional to $x_1^2$, just as in mean field theory. It turns out that
this is true only in high spatial dimensions, namely for $d> 10$. A preliminary
study suggests that $\Delta \bar q(r)\sim x_1^{d/2-3}$ in the dimensional
range of $8< d< 10$. In this paper, we have computed $\Delta \bar q(r)$
in leading order, and it proved to be of order $x_1$ in the whole range of
$6< d< 8$. Beside this regular term, far infrared integration provides
a nonanalytic (anomalous) contribution of the form $\Delta \bar q(r)\sim 
x_1^{1-\epsilon/2}$, which is again expected to emerge in higher order too.
Another remarkable finding of the present calculation is that $\Delta \bar q(r)$
depends only on $x_1$, and not on $\bar g$ (note that $d>6$): this may be a
generic property of $q(x)$ for $d<8$.

Finally, we obtained an important result for the family of the (zero-momentum) replicon
mass, namely
\[
\Gamma_{\text{repl}}(x;u,v)=x_1\,(wq_1)\,\big[
(r_1^2+r_2^2)/2-r^2\big];\qquad
0\le r=x/x_1\le r_1=u/x_1,\,r_2=v/x_1\le 1. 
\]
This formula is in complete agreement with that of the truncated model of
mean field theory \cite{DeKo83,beyond}, any effect of the short range interaction
and the geometry of the hypercubic lattice is embedded in $x_1$ and $q_1$.
Stability of the ultrametric RSB phase below six dimensions is thus demonstrated
along the same lines as in mean field theory.

As an important task in the near future, one should reanalyze the small
momentum behavior of $G_{11}^{xx}(\mathbf p)$ (this is the object studied
numerically in three dimensions in Refs.\ \cite{Contucci_et_al,%
Janus_2010_1,Janus_2010_2}), and finding out
how it changes when crossing the upper critical dimension. Now this seems to be
feasible by the knowledge collected for the first order self-energy in
Appendix \ref{B}.

\begin{acknowledgments}
Useful comments and suggestions from Imre Kondor and Giorgio Parisi are
highly appreciated. I am also very grateful to Imre Kondor for his thorough review of
the manuscript prior to publication.
\end{acknowledgments}


\appendix

\section{Some results for the free propagators in the near and far
infrared regimes}
\label{A}

Equations like (\ref{G_sigma_G}) and (\ref{Sigma1}) to (\ref{Sigma7})
are complicated but manageable in the large mass regime where everything can be
expressed in terms of the two propagators introduced in (\ref{R_L}).
Here the relevant propagator components\footnote{More precisely, their leading terms.}
are listed in dimensionless form,
and, for easing the notation, we keep the old notations for the dimensionless
quantities: i.e.\ $(2wq_1)\, G\rightarrow G$ and $p^2/(2wq_1)
\rightarrow p^2$. All the results below are taken from Ref.\ \cite{beyond},
see also (\ref{R_L_etc}):
\begin{equation}\label{propagator_list}
\begin{gathered}
G^{xx}_{R}=G_R,\qquad G^{xx}_{L}=G_L\big[1-(1-r^2)\,G_R^2\big],\qquad G^{xx}_{LA}=
G_L^2\big[1+\frac{3}{2}(1-r^2)\,G_R-\frac{1}{2}(1-r^2)^2\,G_R^3\big],\\[5pt]
G^{xx_1}_{\,\,\,1}=\frac{1}{2}r\,G_RG_L\big[(1+2G_L)+(1-r^2)\,G_RG_L\big],
\qquad \delta G^{xx_1}=\frac{1}{2}r\,G_RG_L,\qquad G^{x_1x_1}_{\,\,\,x}=
r^2\,G_RG_L^2.
\end{gathered}
\end{equation}

As an application of these formulae, the leading contribution of the
replicon self-energy (\ref{Sigma1}) at zero momentum is easily derived
as
\begin{equation}\label{Sigma1_p0}
\Sigma^{xx}_{R}(\mathbf p=\mathbf 0)=\delta M+x_1\,wq_1\,r^2
\end{equation}
where the integrals obtained were substituted by the results in Eqs.\
(\ref{x1_leading}) and (\ref{deltaM}).

We can also use the leading large mass propagators in (\ref{G_sigma_G}),
and after inserting the expressions from (\ref{propagator_list}), we get
the following result valid in the near-infrared ($p^2\sim 2wq_1$) regime:
\begin{multline}\label{G_sigma_G_v}
(2wq_1)^2\,(G\Sigma G-\delta M\,G^2)^{xx_1}_{\,\,\, 1}=\\[4pt]
\frac{1}{2}r\,\big[ (2G_R^3+G_R^2G_L^2)-2r^2G_R^3G_L^2\big]\,
(\Sigma^{xx}_{R}-\delta M)-\frac{1}{2}r\,G_R^2G_L\,(\Sigma^{xx}_{L}-\delta M)
-\frac{1}{2}r\,G_RG_L^2\,\Sigma^{x_1x_1}_{\,\,\,x}
+G_RG_L\,\Sigma^{xx_1}_{\,\,\,1}\\[4pt]+\big[ (2G_RG_L^2+G_R^2G_L^2)
-3r^2\,G_R^2G_L^2\big]\,\delta \Sigma^{xx_1}
+\frac{1}{2}r\,\big[(G_RG_L^2+3G_R^2G_L^2-2G_R^2G_L^3)-2r^2\,G_R^2G_L^3\big]\,
(\Sigma^{x_1x_1}_{L}-\delta M)\\[4pt]
+\frac{1}{2}r\,G_RG_L^2\,\Sigma^{x_1x_1}_{LA}.
\end{multline}

The free propagator components in the far infrared region (small mass regime)
have the following leading term (restoring now the dimensional dependence of
$p^2$ again):
\begin{equation}\label{G_small_mass}
G\sim \frac{1}{x_1^k\,p^2}\,\,G^{(s)}\left[\frac{p^2}{x_1\,(2wq_1)}\right],
\end{equation}
and using the matching condition (see Sec.\ \ref{free_propagator}) between
the near and far infrared regimes, one can infer the $k$ exponent from the
leading infrared power of the large mass form in (\ref{propagator_list}):
\[
G\sim \frac{1}{p^{2(1+k)}}.
\]
The following classes are found:
\begin{itemize}
\item $k=2$: $G^{xx}_{LA}$, $x<x_1$.
This is the most infrared divergent propagator of all.
\item $k=1$: $G^{xx}_{L}$ and $G^{xx_1}_{\,\,\,1}$, $x<x_1$.
\item $k=0$: $G^{xx}_{R}$, $G^{x_1x_1}_{\,\,\,1}$, $\delta G^{xx_1}$,
and $G^{x_1x_1}_{\,\,\,x}$, $x\le x_1$.
\item $k=-1$: $G^{x_1x_1}_{L}$ and $G^{x_1x_1}_{LA}$.
\end{itemize}

\section{The self-energy components appearing in Eq.\ (\ref{G_sigma_G})}
\label{B}

A generic component of the one-loop self-energy matrix is shown in
(\ref{sigma}). In the present calculation, we can use the SBA when computing
the self-energy components occurring
in (\ref{G_sigma_G}). These components are linear combinations according
to the rules in Eq.\ (\ref{R_L_etc}), now applied for the self-energy matrix.
For easing the notation, the momentum arguments are
not displayed: as a general rule, the first $G$ is always at momentum
$\mathbf q$, whereas the second one in a product is at $\mathbf {p-q}$.
The interchange
$\{\mathbf q  \leftrightarrow  \mathbf p-\mathbf q\}$ means, just as in
(\ref{sigma}), the same terms but with interchanging $\mathbf q$ and
$\mathbf {p-q}$. After some replica algebra one obtains:
\begin{align}\label{Sigma1}\allowdisplaybreaks
&\Sigma^{xx}_{R}(\mathbf p)=w^2\,\frac{1}{N}\sum_{\mathbf q}
\big[G^{xx}_{R}(G^{x_1x_1}_{\,\,\,x}-2G^{x_1x_1}_{L}-G^{x_1x_1}_{LA})
+G^{xx}_{L}G^{x_1x_1}_{L}+
4\,\delta G^{xx_1}\delta G^{xx_1}
+\{\mathbf q  \leftrightarrow  \mathbf p-\mathbf q\}\big],\\[3pt]
&\Sigma^{xx}_{L}(\mathbf p)=w^2\,\frac{1}{N}\sum_{\mathbf q}
\big[G^{xx}_{R}(2G^{x_1x_1}_{\,\,\,x}+
3G^{x_1x_1}_{R}-3G^{x_1x_1}_{L}-2G^{x_1x_1}_{LA})+
G^{xx}_{L}(G^{x_1x_1}_{\,\,\,x}-G^{x_1x_1}_{L}-G^{x_1x_1}_{LA})\notag \\
&\qquad\qquad\qquad\qquad\quad -2G^{xx}_{LA}G^{x_1x_1}_{L}
-8G^{xx_1}_{\,\,\,1}\delta G^{xx_1}+16\,\delta G^{xx_1}\delta G^{xx_1}
+\{\mathbf q  \leftrightarrow  \mathbf p-\mathbf q\}
\big],\label{Sigma2}\\[3pt]
&\Sigma^{x_1x_1}_{\,\,\,x}(\mathbf p)=\frac{1}{2}w^2\,\frac{1}{N}\sum_{\mathbf q}
\big[G^{xx}_{R}(5G^{xx}_{R}-6G^{xx}_{L}-4G^{xx}_{LA})+G^{xx}_{L}G^{xx}_{L}
+8G^{x_1x_1}_{\,\,\,x}G^{x_1x_1}_{\,\,\,x}\notag\\
&\qquad\qquad\qquad\qquad\qquad\,\,-32G^{xx_1}_{\,\,\,1}\delta G^{xx_1}
+24\,\delta G^{xx_1}\delta G^{xx_1}
+\{\mathbf q  \leftrightarrow  \mathbf p-\mathbf q\}
\big],\label{Sigma3}\\[3pt]
&\Sigma^{xx_1}_{\,\,\,1}(\mathbf p)=w^2\,\frac{1}{N}\sum_{\mathbf q}
\big[
(-G^{xx}_{R}+G^{xx}_{L}+2G^{x_1x_1}_{\,\,\,x}-G^{x_1x_1}_{L}
-2G^{x_1x_1}_{LA})\,G^{xx_1}_{\,\,\,1}\notag\\&
\qquad\qquad
+(-2G^{xx}_{R}+2G^{xx}_{L}+2G^{xx}_{LA}-G^{x_1x_1}_{\,\,\,x}-6G^{x_1x_1}_{R}+
6G^{x_1x_1}_{L}+5G^{x_1x_1}_{LA})\,\delta G^{xx_1}
+\{\mathbf q  \leftrightarrow  \mathbf p-\mathbf q\}
\big],\label{Sigma4}\\[3pt]
&\delta \Sigma^{xx_1}(\mathbf p)=w^2\,\frac{1}{N}\sum_{\mathbf q}
\big[(-G^{xx}_{R}-2G^{x_1x_1}_{L})\,G^{xx_1}_{\,\,\,1}\notag\\
&\,\,\quad\qquad\qquad\qquad\qquad\,\,
+(G^{xx}_{L}+2G^{x_1x_1}_{\,\,\,x}+G^{x_1x_1}_{L}
-2G^{x_1x_1}_{LA})\,\delta G^{xx_1}
+\{\mathbf q  \leftrightarrow  \mathbf p-\mathbf q\}
\big],\label{Sigma5}
\\[3pt]
&\Sigma^{x_1x_1}_{L}(\mathbf p)=w^2\,\frac{1}{N}\sum_{\mathbf q}
\big[
3G^{x_1x_1}_{R}G^{x_1x_1}_{R}-4G^{x_1x_1}_{L}G^{x_1x_1}_{L}
-8G^{x_1x_1}_{L}G^{x_1x_1}_{LA}
+\{\mathbf q  \leftrightarrow  \mathbf p-\mathbf q\}
\big],\label{Sigma6}\displaybreak\\[3pt]
\notag &\Sigma^{x_1x_1}_{LA}(\mathbf p)=
\frac{1}{2}w^2\,\frac{1}{N}\sum_{\mathbf q}
\big[3G^{x_1x_1}_{R}G^{x_1x_1}_{R}+21G^{x_1x_1}_{L}G^{x_1x_1}_{L}
-24G^{x_1x_1}_{R}G^{x_1x_1}_{L}
-8G^{x_1x_1}_{LA}G^{x_1x_1}_{LA}\\
&\,\qquad\qquad\qquad\qquad\qquad\,\,
+16G^{x_1x_1}_{L}G^{x_1x_1}_{LA}
+\{\mathbf q  \leftrightarrow  \mathbf p-\mathbf q\}
\big].\label{Sigma7}
\end{align}

Any of the self-energy components above has the following generic structure:
\begin{equation}\label{sigma_generic}
\Sigma(\mathbf p)=x_1(wq_1)\big[
\tilde C_d\,\bar \Lambda^{2-\epsilon}+f(p^2/2wq_1)\,\bar\Lambda^{-\epsilon}
+\sigma^{(a)}(p^2/2wq_1)+\sigma^{(na)}
\big]
\end{equation}
where $\sigma^{(a)}$ and $\sigma^{(na)}$ are analytical and nonanalytical
contributions in $x_1$, respectively, and both have corrections which are
smaller by factors of $x_1$ (or higher integer powers of $x_1$)
and, therefore, are irrelevant for the present calculation.
While $f$ and $\sigma^{(a)}$ have simple one-mass-scale momentum dependence,
$\sigma^{(na)}$ has the double-mass-scale structure like the free propagator:
\begin{equation}\label{sigma_na}
\sigma^{(na)}(p^2)=x_1^{-\frac{\epsilon}{2}+a}
\begin{cases}
f^{(l)}(p^2/2wq_1) & \text{for $p^2\sim 2wq_1$}\\[4pt]
x_1^{-b}\,f^{(s)}(p^2/x_1\,2wq_1) & \text{for $p^2\sim x_12wq_1$}.
\end{cases}
\end{equation}
$\tilde C_d$ is always zero except for $\Sigma^{xx}_{R}$, $\Sigma^{xx}_{L}$,
and $\Sigma^{x_1x_1}_{L}$; in these cases $\tilde C_d=-4C_d^{-1}/(d-4)$ where
$C_d\equiv 6\,\Gamma(d/2-2)\Gamma(4-d/2)$. Some properties of the self-energy
components which are relevant for the present calculation are summarized
in Table \ref{table_sigma}.
\begin{table}
\caption{The most relevant properties of the self-energy components. The
functions $f$, $\sigma^{(a)}$, and $f^{(l)}$ --- their argument being the
dimensionless momentum squared, i.e.\ $p^2/(2wq_1)\rightarrow p^2$ ---,
and the exponents $a$, $b$ are defined in Eqs.\ (\ref{sigma_generic}) and
(\ref{sigma_na}). While $f(p^2)$ is exact, only the leading $1/\epsilon$
term for $\sigma^{(a)}$ and $f^{(l)}$ is shown ($\epsilon=6-d$).
$C_d$ is the notation for $6\,\Gamma(d/2-2)\Gamma(4-d/2)$, whereas
$r\equiv x/x_1$ throughout the paper.
}
\label{table_sigma}
\[
\begin{array}{|c|c|c|c|c|c|}
\hline
 & f(p^2) & \sigma^{(a)}(p^2) & f^{(l)}(p^2) &
a & b\\
\hline\hline
\Sigma^{xx}_{R} & 4C_d^{-1}\big(1+\frac{4-d}{d}\,p^2\big)\,\frac{1}{\epsilon} &
\big(\frac{2}{9}\,p^2-\frac{2}{3}\big)\,\frac{1}{\epsilon}+O(1)
& O(1) & 2 & 1 \\
\hline
\Sigma^{xx}_{L} & 4C_d^{-1}\,\frac{4-d}{d}\,\,p^2\,\frac{1}{\epsilon} &
\big[\frac{2}{9}\,p^2-\frac{2}{3}(1-r^2)^2\,\frac{1}{p^2+1}\big]
\,\frac{1}{\epsilon}+O(1) &\frac{2}{3}(1-r^2)^2\,\frac{1}{p^2+1}
\,\frac{1}{\epsilon}+O(1)  & 0 &0  \\
\hline
\Sigma^{x_1x_1}_{\,\,\,x} & 0 & -\frac{2}{3}(1-r^2)^2\,\frac{1}{p^2}
\,\frac{1}{\epsilon}+O(1)
&\frac{2}{3}(1-r^2)^2\,\frac{1}{p^2}\,\frac{1}{\epsilon}+O(1) & 0 & 1 \\
\hline
\Sigma^{xx_1}_{\,\,\,1} & 2C_d^{-1}\,r\,\frac{1}{\epsilon}&
-\frac{1}{3}\,r\,\big[1-(1-r^2)^2\,\frac{1}{p^2(p^2+1)} \big]
\,\frac{1}{\epsilon}+O(1) & -\frac{1}{3}\,r(1-r^2)^2\,\frac{1}{p^2(p^2+1)}
\,\frac{1}{\epsilon}+O(1)
 &0 &1 \\
\hline
\delta \Sigma^{xx_1} & 2C_d^{-1}\,r\,\frac{1}{\epsilon}&
-\frac{1}{3}\,r\,\frac{1}{\epsilon}+O(1)& O(1) & 1 & 1 \\
\hline
\Sigma^{x_1x_1}_{L} & 4C_d^{-1}\,\frac{4-d}{d}\,\,p^2\,\frac{1}{\epsilon} &
\frac{2}{9}\,p^2\,\frac{1}{\epsilon}+O(1) & O(1) & 2 & 1\\
\hline
\Sigma^{x_1x_1}_{LA} & 4C_d^{-1}\,\frac{1}{\epsilon}&
-\frac{2}{3}\,\frac{1}{\epsilon}+O(1) & O(1) & 2 & 1\\
\hline
\end{array}
\]
\end{table}

\section{Details of the different self-energy contributions to $X_2$}
\label{C}

$X_2$ of Eq.\ (\ref{X_2}) can be studied, and its relevant terms computed,
using Eqs.\ (\ref{m_c}), (\ref{deltaM}), (\ref{G_sigma_G}),
and the results of the preceding
appendices, mainly (\ref{propagator_list}), (\ref{G_sigma_G_v}), 
(\ref{Sigma1}) to (\ref{Sigma7}), and Table \ref{table_sigma}. 
\begin{list}{$\bullet$}{\setlength\leftmargin{0pt}}
\item $\Sigma^{xx}_{R}$:

This replicon contribution can be conveniently evaluated by using
(\ref{Sigma1_p0}) and writing $\Sigma^{xx}_{R}(\mathbf p)-\delta M=
\big[\Sigma^{xx}_{R}(\mathbf p)-\Sigma^{xx}_{R}(\mathbf p=\mathbf 0)\big]
+x_1\,wq_1\,r^2$. It is useful to add $-2m_c^{(2)}q_1r$ with the first term
in $m_c^{(2)}$, see (\ref{m_c}), to the contribution with the zero-momentum
subtraction to yield:
\begin{description}
\item[(i)] $\Sigma^{xx}_{R}(\mathbf p)-\Sigma^{xx}_{R}(\mathbf p=\mathbf 0)$ term:
\begin{multline*}
X_2=-2C_d^{-1}\,r\,(wq_1^2)\,x_1^2\,\int_0^{\bar \Lambda}dp\, p^{d-1}\\
\times\Big\{
2G_R^3\,\Big[\sigma^{(a)}(p^2)+\frac{d-4}{3\epsilon}-r^2+\frac{4}{4-d}\,C_d^{-1}\,
\frac{\Gamma(\frac{d}{2})\Gamma(\frac{d}{2}-1)^2\Gamma(3-\frac{d}{2})}{\Gamma(d-2)}
\,p^{2-\epsilon}\Big]\\
+(G_R^2G_L^2-2r^2\,G_R^3G_L^2)\,
\Big[\sigma^{(a)}(p^2)+\frac{d-4}{3\epsilon}-r^2-\frac{4}{d}\,C_d^{-1}\,
\frac{d-4}{\epsilon}\,\bar\Lambda^{-\epsilon}\,p^2\Big]\Big\},
\end{multline*}
and $X_2$ takes the form for generic $6<d<8$:
\begin{equation}\label{minta}
X_2=(wq_1^2)\,x_1^2\,(A_d\,\bar\Lambda^{-2\epsilon}+A'_d\,\bar\Lambda^{-\epsilon}
+A''_d)
\end{equation}
with some dimension- and $r$-dependent amplitudes $A_d$, $A'_d$,
and $A''_d$.\footnote{%
See (\ref{sigma_generic}) and Table \ref{table_sigma} for notations and results.
For the sake of avoiding complicated notations, we will not indicate the actual
self-energy component in the quantities like $\sigma^{(a)}$, $A_d$, $A'_d$,
etc., although they are different for different self-energy
components listed here.} In fact $A_d\sim r$, making the second derivative
of $X_2$ proportional to $\bar\Lambda^{-\epsilon}$:
\[
\frac{d^2}{dr^2}X_2=(wq_1^2)\,x_1^2\,\frac{d^2}{dr^2}
(A'_d\,\bar\Lambda^{-\epsilon}+A''_d).
\]

The integral is valid even in six dimensions where we get:
\[
X_2=(wq_1^2)\,x_1^2\,\left(-\frac{7}{27}\,r\,\ln^2 \bar\Lambda
+\frac{49}{162}\,r\,\ln \bar\Lambda + \frac{20}{27}\,r^3\,
\ln \bar\Lambda+O(1)\right).
\]

\item[(ii)] The $x_1( wq_1) r^2$ term:

Inserting this term into (\ref{G_sigma_G_v}), and after some manipulations
with the propagators of Eq.\ (\ref{propagator_list}), it follows:
\begin{align*}
X_2&=-2C_d^{-1}\,r^3\,(wq_1^2)\,x_1^2\,\int_0^{\bar \Lambda}dp\, p^{d-1}
G_RG_L\big[2G_L+5G_RG_L+2(1-r^2)G_R^2G_L\big]\\[4pt]
&=(wq_1^2)\,x_1^2\,
(A'_d\,\bar\Lambda^{-\epsilon}+A''_d).
\end{align*}

The classification of the far infrared propagators in the end of Appendix
\ref{A} makes it possible to compute the contribution of the small mass regime
$p^2\sim x_1(2wq_1)$. One gets a dangerous term nonanalytical in $x_1$,
namely $X_2\sim (wq_1^2)\,x_1^{2-\epsilon/2}$, which yields a $\ln x_1$
in six dimensions:
\[
X_2=(wq_1^2)\,x_1^2\,\left[-\frac{2}{3}\,r^3\,\ln \bar\Lambda+
\frac{1}{3}\,r^3(1-r^2)\,\ln x_1+O(1)\right],\qquad\quad d=6.
\]
\end{description}
Adding together the results of (i) and (ii), one finally gets the 
$\Sigma^{xx}_{R}$-insertion result in six dimensions:
\begin{equation}\label{X_2^1}
X_2=(wq_1^2)\,x_1^2\,\left[-\frac{7}{27}\,r\,\ln^2 \bar\Lambda
+\frac{49}{162}\,r\,\ln \bar\Lambda+\frac{2}{27}\,r^3\,\ln \bar\Lambda
+\frac{1}{3}\,r^3(1-r^2)\,\ln x_1+O(1)\right].
\end{equation}

\item $\Sigma^{xx}_{L}$ and $\Sigma^{x_1x_1}_{L}$:

The two longitudinal terms in the right hand side of Eq.\ (\ref{G_sigma_G_v})
can be most conveniently written as
\[
-\frac{1}{2}r\,G_R^2G_L\,\big(\Sigma^{xx}_{L}-\Sigma^{x_1x_1}_{L}\big)
+r\,G_R^2G_L^2\big[(1-G_L)-r^2\,G_L\big]\,\big(\Sigma^{x_1x_1}_{L}-\delta M\big).
\]
\begin{description}
\item[(i)] The $\Sigma^{xx}_{L}-\Sigma^{x_1x_1}_{L}$ part produces, due to
the $\sigma^{(na)}$ in $\Sigma^{xx}_{L}$, a dangerous nonanalytical term
in $X_2$, but the $\bar\Lambda^{-2\epsilon}$ and $\bar\Lambda^{-\epsilon}$
contributions are canceled by the subtraction:
\[
X_2=(wq_1^2)\,x_1^2\,(A''_d+B_d\,x_1^{-\epsilon/2}).
\]
The interplay between the analytical and nonanalytical terms in $x_1$
[see Eqs.\ (\ref{sigma_generic}), (\ref{sigma_na}), and the entries for
$\Sigma^{xx}_{L}$ in Table \ref{table_sigma}] produces the $\ln x_1$ for
$d=6$:
\begin{align}
X_2&=-2 w\,(2wq_1)^{1-\epsilon/2}
\,\frac{1}{N}\sum_{\mathbf p}\left[
-\frac{1}{2}r\,G_R^2G_L\right]\,\times\left[-
x_1(wq_1)\frac{2}{3}(1-r^2)^2\,G_L\big(1-x_1^{-\epsilon/2}\big)
\frac{1}{\epsilon}\right]\notag\\[6pt]\label{IR}
&\overset{d=6}{=}(wq_1^2)\,x_1^2\,
\left[-\frac{1}{18}\,r(1-r^2)^2\,\ln x_1+O(1)\right].
\end{align}
\item[(ii)] The $\Sigma^{x_1x_1}_{L}-\delta M$ insertion 
yields the contribution to $X_2$ in a generic dimension 
just as in (\ref{minta}). In six dimensions, it becomes:
\[
X_2=(wq_1^2)\,x_1^2\,\left(-\frac{2}{27}\,r\,\ln^2 \bar\Lambda
-\frac{11}{81}\,r\,\ln \bar\Lambda + \frac{4}{27}\,r^3\,
\ln \bar\Lambda+O(1)\right).
\]
\end{description}
Finally the complete six-dimensional result for the 
$\Sigma^{xx}_{L}$ and $\Sigma^{x_1x_1}_{L}$ insertions is the sum of
(i) and (ii):
\begin{equation}\label{X_2^2}
X_2=(wq_1^2)\,x_1^2\,\left[-\frac{2}{27}\,r\,\ln^2 \bar\Lambda
-\frac{11}{81}\,r\,\ln \bar\Lambda+\frac{4}{27}\,r^3\,\ln \bar\Lambda
-\frac{1}{18}\,r(1-r^2)^2\,\ln x_1+O(1)\right].
\end{equation}

\item $\Sigma^{x_1x_1}_{\,\,\,x}$:

This self-energy is ultraviolet convergent, which is reflected by the
fact that $f(p^2)\equiv 0$, see Table \ref{table_sigma}. In fact, the whole
two-loop graph built up from this self-energy is finite for $\Lambda\to
\infty$, and there is no $\ln \bar\Lambda$ in six dimensions. The leading
infrared contribution is, however, exactly the same as in the case of
the longitudinal self-energy, i.e. (\ref{IR}): see Eq.\ (\ref{G_sigma_G_v})
and the entries in Table \ref{table_sigma}. We thus finally have:
\begin{equation}\label{X_2^3}
X_2=(wq_1^2)\,x_1^2\,\left[
-\frac{1}{18}\,r(1-r^2)^2\,\ln x_1+O(1)\right],\qquad\quad d=6.
\end{equation}

\item $\Sigma^{xx_1}_{\,\,\,1}$:

This term --- which is somewhat complicated, but 
manageable when we are looking for the logarithms in six dimensions
--- must be treated together with the
second part of the $-2m_c^{(2)}q_1r$ subtraction, see Eq.\ (\ref{m_c}).
In generic dimensions
$d$ it has the structure of Eq.\ (\ref{minta}) together with a
dangerous nonanalytical contribution $X_2\sim (wq_1^2)\,x_1^{2-\epsilon/2}$.
Although the $\bar\Lambda^{-2\epsilon}$ term suggests that a $\ln^2 \bar\Lambda$
should exist in $d=6$, the two such terms cancel out each other. Finally we have:
\begin{equation}\label{X_2^4}
X_2=(wq_1^2)\,x_1^2\,\left[
\frac{5}{9}\,r\,\ln \bar\Lambda-\frac{1}{9}\,r^3\,\ln \bar\Lambda
-\frac{1}{18}\,r(1-r^2)^2\,\ln x_1+O(1)\right],\qquad\quad d=6.
\end{equation}

\item $\delta \Sigma^{xx_1}$:

For generic $d$, $X_2$ takes the form of (\ref{minta}), and there is no dangerous
nonanalytic correction. One can relatively easily find:
\begin{equation}\label{X_2^5}
X_2=(wq_1^2)\,x_1^2\,\left[
\frac{2}{9}\,r\,\ln^2 \bar\Lambda
-\frac{1}{3}\,r^3\,\ln \bar\Lambda+O(1)\right],\qquad\quad d=6.
\end{equation}

\item $\Sigma^{x_1x_1}_{LA}$:

We have again an $X_2$ like in Eq.\ (\ref{minta}) without any dangerous
nonanalytic correction. The six-dimensional limit yields
\begin{equation}\label{X_2^6}
X_2=(wq_1^2)\,x_1^2\,\left[
\frac{1}{9}\,r\,\ln^2 \bar\Lambda
-\frac{1}{18}\,r\,\ln \bar\Lambda+O(1)\right],\qquad\quad d=6.
\end{equation}
\end{list}


\end{document}